# Enhanced D-D Fusion Rates when the Coulomb Barrier Is Lowered by Electrons

Alfred Y. Wong[a]*, Alexander Gunn[a], Allan X. Chen[a,b], Chun-Ching Shih[c], Mason J. Guffey[a]

**May 20, 2021**

[a]   (Alpha Ring International Limited: Monterey, CA, USA)
[b]   (Adelphi Technologies, Inc: Redwood City, CA, USA)
[c]   (Chun-Ching Shih: CA, USA)
*    Corresponding Author: awong@alpharing.com

**Abstract**

A profusion of unbound, low-energy electrons creates a local electric field that reduces Coulomb potential and increases quantum tunneling probability for pairs of nuclei. Neutral beam-target experiments on deuterium-deuterium fusion reactions, observed with neutron detectors, show percentage increases in fusion products are consistent with electron-screening predictions from Schrödinger wave mechanics. Experiments performed confirm that observed fusion rate enhancement with a negatively biased target is primarily due to changes to the fusion cross section, rather than simply acceleration due to electrostatic forces.

# 1. Introduction

The effect of electrons on fusion reaction rates has been investigated for many years to understand its influence on stellar nucleosynthesis. The screening effects due to bound electrons have been studied for various nuclear reactions, including D-D, D-$^3$He, $^3$He-$^3$He, p-$^{11}$B, and p-$^7$Li [1–6]. For the reaction p-$^{11}$B, the screened cross section has been measured for energies between 17 and 134 keV [1]. In this range, the screening potential is only 0.25 to 2 percent of the total interaction energy, resulting in enhancement of fusion cross sections of less than 10 percent [5]. Although the bound electrons produce relatively high electric fields, the cumulative effect is largely cancelled out by the positive charge of the nucleus, thus the net effect on the effective cross section is not significant.

The screening effect due to thermal electrons, such as in two-component plasmas or neutral metals, has also been studied extensively [5–14]. In an overall neutral system, the positively charged ions are immersed in a sea of free electrons, which tend to cluster around regions near individual ions, at the Debye-Hückel radius, having a charge neutralization effect similar to that of orbital electrons around a nucleus. For metals, this screening distance is usually in the range of nanometers. The effects of the thermal electrons on nuclear reactions have been measured with a deuterium target, embedded in metals [9]. This magnitude of the thermal-electron screening is close to the effect by bound electrons, around tens to hundreds of eV. The enhancement of cross sections is also limited to only a few percent because the screening fields are not coherent [15].

The combined screening effects of bound and thermal electrons can be used to explain the minor change of nuclear reaction rates in stars [16]. To produce fusion energy more effectively in a laboratory setting, reaction rates must be raised by orders of magnitude compared to the processes of stellar fusion. This can only be achieved with more aggressive screening [17].

Noting that both bound and thermal electrons produce measurable enhancement in fusion cross sections, we wish to report on a process of using a profusion of low-energy, free electrons in target materials to generate screening fields which may be able to reduce the Coulomb barrier significantly, such that fusion cross sections are improved from vanishingly small to a level of interest (~ $10^{-30}$ to $10^{-40}$ m$^2$) for significant reaction rates. This process involves ions as well as high-density neutrals. The presence of neutrals in the system has the advantage of yielding high beam densities and significant fusion events without causing plasma instabilities due to space charge.

A presentation of a Schrödinger wave mechanics approach to predict the effects of screening fields is given in Section 2.1 below. An experiment designed to use low energy neutral beams (center-of-mass energy of 25 keV) of deuterium to interact with a target of high-density deuterium (biased up to –20 keV) demonstrates fusion through detection of energetic neutrons, $^3$He, Tritium and protons. An increased fusion rate indicates the efficiency of shielding by free electrons. The tunneling probability through the screened Coulomb barrier and the associated fusion cross sections for D-D reactions is shown in Section 2.2. Alternative possible explanations of fusion rate enhancement are considered in Section 5.

## 2. Theory/Calculations

### 2.1 Schrödinger Equation and Screening Potentials

$$-\frac{\hbar^2}{2m}\nabla^2\psi + \left[\left(\frac{Z_1 Z_2 e^2}{4\pi\epsilon r} - U_s\right) - E\right]\psi = 0 \qquad (1)$$

The Schrödinger equation in Eq. 1 includes the screening potential $U_s$ to represent the electron screening effect as a reduction of the Coulomb barrier between two nuclei of $Z_1$ and $Z_2$ within the range where quantum wave properties of the particles are not negligible. At these interaction distances the screening potential can be treated as a constant reduction of the internuclear Coulomb potential. In such a case, Eq. 1 can be readily used to derive screened fusion cross sections for a given beam energy.

We now consider the target as a cylinder with radius $R_1$ located coaxially within a grounded cylinder of radius $R_2$, perpendicular to the beam path. When a target is negatively biased the field and potential energy in the space between $R_1$ and $R_2$ can be found as:

$$E(R) = \frac{V}{R \ln\left(\frac{R_2}{R_1}\right)} \qquad (2)$$

$$U(R) = -eZ_1 \frac{\ln\left(\frac{R_2}{R}\right)}{\ln\left(\frac{R_2}{R_1}\right)} V \qquad (3)$$

Here $R$ is the initial radial position of the particle, $V$ is the bias voltage, and $eZ_1$ is the charge of the projectile. On the surface of a 0.5 cm rod biased at –20 kV in a 1.5 cm tube, the electric field $E_1$ at the rod surface without any plasma in the medium is calculated to be approximately 4 MV/m.

During our experiments, a plasma was formed in the vicinity of the target electrode. The ion density in the sheath region creates a positive potential that modifies the electric field at the surface of the target electrode. The net screening energy $U_s$ experienced by the impacting beam particles and the target must be computed from the imposed external fields and the screening from plasmas. The Debye length has been computed to be $10^{-3}$ cm.

### 2.2 Tunneling Probability and Fusion Cross Sections

A nucleus approaching another nucleus on the target surface experiences a generally repulsive force as expected. It also experiences the collective, attractive force of the electrons grouped on the biased target. The wave nature of the incoming beam nucleus allows it to propagate through the barrier as an evanescent wave. The probability of its penetration, or tunneling, depends on its velocity and the amount by which the Coulomb barrier is lowered. Fusion cross section is related to this tunneling probability.

The probability of the particle tunneling through a 1-D barrier can be obtained using a simple wave approach and the result is (derived in Appendix A):

$$P = exp\left[-\frac{2\sqrt{2m}}{\hbar}\int_{r_1}^{r_2} \sqrt{U(r)-E}\,dr\right] \quad (4)$$

Here $r_2$ is the nuclear radius and $r_1$ is the zero-velocity radius. For the modified Coulomb barrier, the potential $U$ and the energy $E$ are given as

$$U = \frac{Z_1 Z_2 e^2}{4\pi\epsilon r} - U_S \quad ; \quad E = \frac{Z_1 Z_2 e^2}{4\pi\epsilon r_1} \quad (5)$$

Here $U_s$ is the screening potential energy is assumed to be equal to the applied bias voltage ($V_b$) multiplied by a shielding factor $\eta$ where $U_s = \eta eV_b$. This factor $\eta$ depends on the characteristics of plasmas produced around this target. The free-space potential profile is modified by the density and temperature of plasma in front of the target. Our present experiment is designed to verify only the functional dependence of the shielding as shown in Eq. 7 below. The absolute value $\eta$ is to be determined experimentally.

Using the WKB approximation,

$$P \approx exp\left(-\sqrt{\frac{E_G}{E}}\right) \quad (6)$$

where $E_G$ is the Gamow energy $E_G = (\pi\alpha Z_1 Z_2)^2 2m_r c^2 = 986\ keV$, $m_r$ is the reduced mass, and the $\alpha$ is the fine-structure constant $\alpha = \frac{e^2}{\hbar c} = \frac{1}{137.04}$.

The fusion cross-section for a 3-D Coulomb potential is (derived in Appendix B):

$$\sigma(E, U_s) \sim \frac{P}{E} = \frac{S(E+U_s)}{E+U_s} exp\left(-\sqrt{\frac{E_G}{E+U_s}}\right) \quad (7)$$

We have included in $\sigma(E)$, the astrophysical factor $S(E)$, which represents the probability of nuclear reaction once the projectile tunnels through the barrier [1–5].

The rate of fusion per unit volume can be calculated as

$$\frac{dn}{dt} = n_b n_t \sigma v \quad (8)$$

where $n$ is the number of fusion events, $n_b$ is the density of D in the beam, $n_t$ is the density of D nuclei in the target, and $v$ is the velocity of the beam.

An external electric potential as a DC bias in our experiments can affect the rate of fusion, *dn/dt* from Eq. 7, in three ways—a change in $n_t$, σ, or $v$. In our experiment we have carefully account for the possibilities of changes to $n_t$ and $v$ causing the percentage increase of the rate of fusion that we observe, therefore, we present a model of enhancement of fusion rate to be compared with theory relating to modification of the fusion cross section, σ.

The D-D fusion reaction has two branches of equal probability:

Neutronic reaction: $\quad D + D \rightarrow {}^3He\ (0.82\ MeV) + n\ (2.45\ MeV)$ (9)

Aneutronic reaction: $\quad D + D \rightarrow T\ (1.01\ MeV) + p\ (3.02\ MeV)$ (10)

A count of any of the four emitted products ($^3$He, n, T, p) from the system is, therefore, a reliable measure of the overall rate of fusion. The neutron measurement is preferable to charged particle detection for routine measurements since much more care is required to establish controls for interfering signals when measuring the charged particles with the silicon detector than for the neutron measurement, and the neutron detector can be located outside of the pressurized chamber. Using an ion-implanted Silicon detector connected to a multi-channel analyzer to output the detected particle energies (MeV) and number of particles (as pulse height), we observed particle energies correlating to helium-3 (~0.5 MeV), tritium (~0.8 MeV), and proton (~3 MeV) as shown in Fig. 1. The departure from tabulated energies is due to collisions between fusion products and the plasma medium.

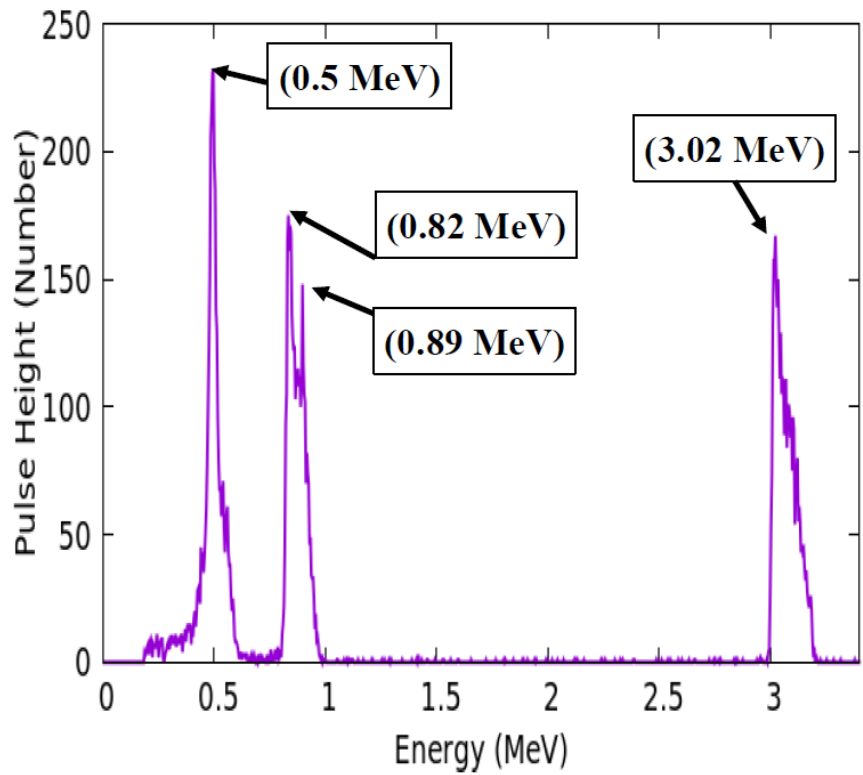

Figure 1: Fusion products (proton, tritium, and helium-3) observed with a silicon detector and multi-channel analyzer from Ortec. The detector was placed within the target chamber with about 5 mTorr deuterium gas, during a D-D fusion experiment a few inches from the target and perpendicular to the beam path. Energies were calibrated with respect to an Am-241 source (calibration data not shown) and are consistent with those listed in Eqs. 9 and 10.

## 3. Material and Methods: Reduction in Coulomb Barrier by Free Electrons

An experiment was designed to quantify the lowering of the Coulomb barrier by electrons using a beam-target interaction configuration. A reproducible, low-power (1.25 W) deuteron beam was chosen to interact with an equally reproducible plasma to allow for digital sampling and signal averaging of many repeated beam and target interactions per second.

The deuteron beam was produced from an ion beam source with a microwave-generated deuterium plasma of density up to $10^{12}$ cm$^{-3}$ with ion species up to 90% [18]. The ion source was floated to high positive potential using an innovative DC blocking method [Unpublished results] which allowed microwave power to be injected into the plasma chamber while holding off the high voltage from ground. As a result, the ion beam produced was able to be sustained at a high energy value of 25 to 100 keV with a grounded experimental chamber.

As the ion beam exits the ion source into the pressurized chamber, the ions traverse through a path of sufficiently high neutral density that the deuterium beam becomes neutralized. The ion source and the experiment chamber were both operated at the same pressure, verified by both convection enhanced Pirani and hot cathode gauges. Based on previous work performed at ORNL [19], the fraction of neutral beam along a drift tube has been analyzed based on published cross-sections of the hydrogen species evolution. These involve all the combinations of H$^+$, H$^0$, H$^-$ and their corresponding molecular species. Based on the calculated value of the line density,

$$x(L) = \int_0^L n(l)dl \tag{11}$$

where $L$ is the total length traversed by the beam and n($l$) is the neutral density as a function of distance, n($l$) being a constant for a given chamber pressure, the ion fractions of the beam can be calculated. The line density was approximately $10^{16}$ cm$^{-2}$ based on the measured values in this system. Referring to the predicted ion fractionation [19] of the hydrogen species, the neutral species fraction in a 40 keV deuterium at the calculated line density is predicted to be at least 80%. The differences between the charge exchange interactions in hydrogen and those of the isotope deuterium are assumed to be negligible.

Deuterons in the beam exit the ion source with a narrow distribution of kinetic energies about $E_i$ and can gain or lose energy through interactions with other deuterons or with electromagnetic fields as they traverse the distance $L$ through the chamber. The beam energy is significantly higher than the average velocity of the deuterium gas at thermal equilibrium. The resultant net momentum of the beam, therefore, remains roughly unchanged in direction by the charge exchange interactions. Deuterium ions will be accelerated toward a negatively biased target, but this interaction is weakest at the beam source, when the ion population in the beam is greatest, and for a given deuteron is limited by the collision frequency. These combined effects result in a deuteron beam that strikes the target consisting of mostly neutral atoms travelling at close to the initial ion beam velocity.

Further neutralization of the beam was accomplished using a parallel set of biased plates along the beam path to set up a perpendicular E-field in order to deflect the residual ion species to minimize the number of ions that strike the target (see Fig. 2). A neutral beam cannot gain energy from a biased target during its transit from source through the chamber. The ion fractionation model used to calculate the collisional neutralization of the initial ion beam also predicts a certain degree of ionization from the continual collisions after the removal of ions using the steering voltage. We expect that the remaining beam that proceeds toward the target is predominantly neutral atoms. The generally neutral character of the beam under these conditions was confirmed by pointing the beam at a 30-degree angle to an axial magnetic field with no observable bending of the beam due to Lorentz forces.

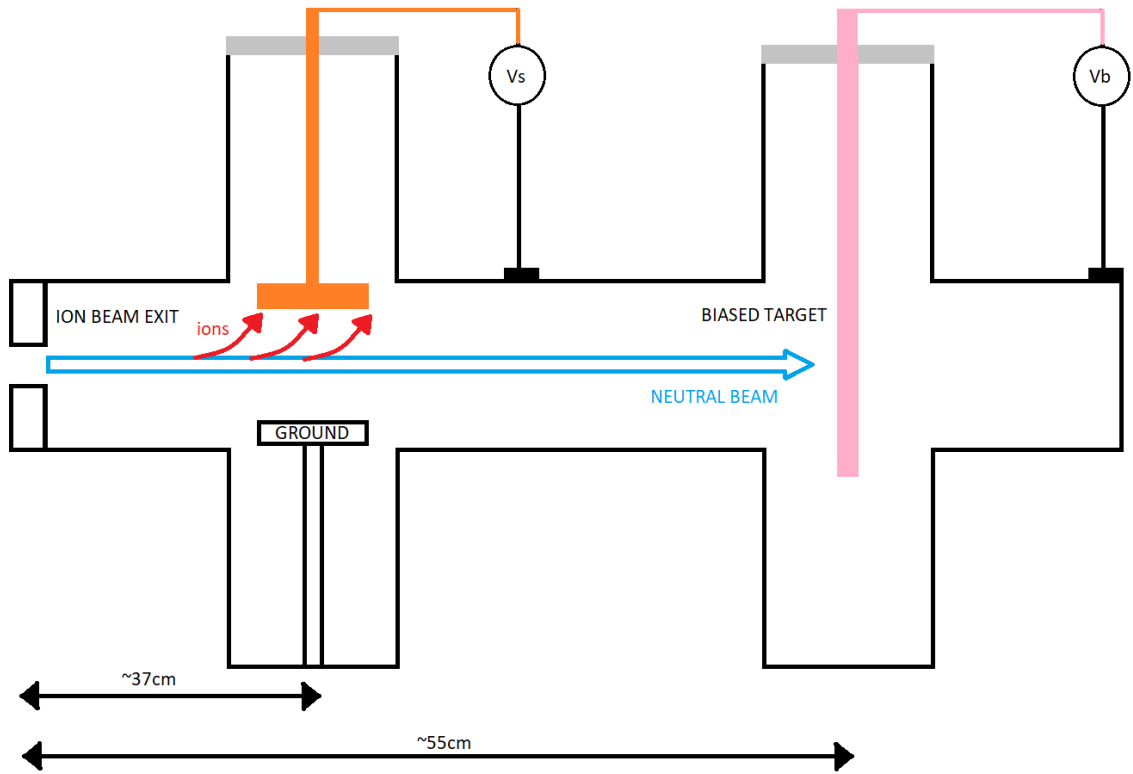

Figure 2: Schematic of beam-target experiment. After exiting the ion source, the particle beam is neutralized as it travels approximately 37 cm to the deflection plates that remove residual ions. The total distance traveled is approximately 55 cm and the typical pressure in the chamber is approximately 6 mTorr as measured by a convection enhanced Pirani gauge. The target is biased between –20 kV to 0 kV.

Ions approaching the target will be accelerated toward the negative bias. The net acceleration is likely to be small for the entire beam, but there will likely be a greater spread in the distribution of kinetic energy. Because of this ionization, some deuterons that would have just passed the target when no bias is applied may get pulled in when the bias potential is present. This change in number of beam particles striking the target in such cases must, therefore, be included in the calculation as a change in $n_b$.

The Debye sheath that forms around the target when a bias potential is applied decreases the effective potential experienced by the beam ions as they approach the target. As the beam nuclei pass through the Debye sheath, however, they will experience an increased local electric potential gradient.

An aluminum rod target of 0.5 cm radius was placed 1.5 cm from the chamber ground and directly in the path of the deuteron beam. The beam was either continuous or pulsed at a constant rate. Loading of deuterons at the target surface, with no bias potential applied, until equilibrium was achieved established a steady baseline fusion rate and maintained the reactant densities constant for each bias and beam energy.

Neutron counts were measured with a proton-recoil fast neutron detector (FND) developed by Alpha Ring (detector design is described in Appendix C). Additional monitoring of the average neutron count rate by a Bonner sphere and a $^3$He thermal neutron detector with a cylindrical polyethylene jacket to slow fast neutrons for detection so that the thermal neutron detector provides information related to both fast and slow neutron counts. The fast neutron detector, however, can also provide information on timing of neutron events in experiments where pulsed particle beams are used.

After beam loading reached equilibrium, neutron counts were recorded for a selected time interval with a bias potential applied to the target to accumulate negative charges. The neutron production rate was measured at each bias value, thus with a different density of free electrons. An increase in neutron production was observed as the target bias potential was varied from 0 to –20 kV. The relative increase in fusion rate is reported here as a percent increase of counts during the measurement period with respect to measured counts in the 0 kV case over the same duration.

**4. Results**

The relative increase in fusion rate with increasing bias potentials of the target, as measured on different days by neutron production, is shown in Fig. 3. The beam power used in these experiments was constant at approximately 1.2 W throughout each data set indicating that the density of ions initially accelerated did not change significantly.

Measurements shown in Fig. 3 were taken with the indicated neutron detectors for 20 minutes at each bias value. Count rate data was collected on a per minute basis. The average 95% confidence interval of the average neutron counts at each bias value, propagated through the calculation of fusion rate enhancement, was calculated to be around ± 17% enhancement.

Because neutron data is obtained at only one location per detector, the comparison with theory is limited to the demonstration of trend variation. Theoretical curves were derived from the assumption that the enhancement in neutron yields is proportional to the percentage increase in cross sections.

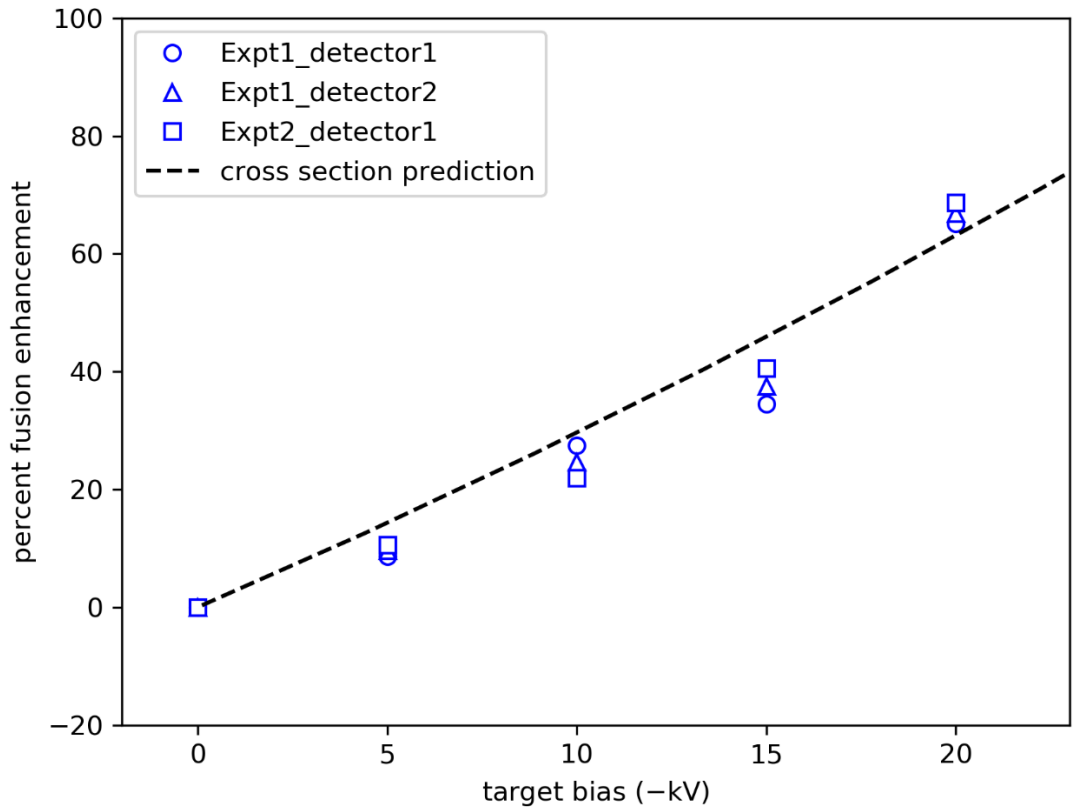

Figure 3: Percent increase in the neutron production rate with the variation of target bias potential. The ion beam at 25 keV and 0.05 mA is generated from a capacitively-coupled plasma by excitation at 2473 MHz, +9 dBm, and 100% duty cycle with -100 V and 0.040 mA electron suppression. Beam neutralization, in a deuterium pressure of 5.3 mTorr for set 1 and 5.4 mTorr for set 2, was further ensured using a -5 kV steering potential perpendicular to the beam path. Each measurement period consists of repeated acquisition of neutron counts over 20 minutes at the specified target bias. The theory curve (dashed line) represents an effective shielding energy calculated using the factor $\eta = 0.22$. Experiment 1 and experiment 2 represent independent data sets from two different runs. Detector 1 represents data from the FND, and detector 2 represents the detection of neutrons that have been slowed by the polyethylene housing of the thermal neutron detectors for the specified data set.

## 5. Discussion

The increased fusion efficiency shown in Fig. 3 is consistent from one data set to another. The relative increase calculated from fast neutron measurements only is also consistent with that observed when neutrons of all energies are measured. The trend in observed enhancement in the rate of fusion for each data set is within measurement uncertainties and distinct from that predicted by an increased system velocity from the additional acceleration expected for ions approaching the biased target.

The small population of ions in the beam at the time the beam encounters the target will be accelerated toward the negative bias. Some of these deuterium ions in the fringes of the beam profile that would otherwise pass the target when no bias is applied may add to the effective reactant density when the bias potential is present. Since we have a mostly neutral beam, and this density increase would come from regions of low deuteron density, it is unlikely that the observed enhancement with increasing bias could be attributed to these additional collisions.

The most significant variable from Eq. 8 that can account for the higher fusion rates is, therefore, an increase in the effective cross section, that is in turn related to Coulomb barrier reduction. The combined effects of charges near the nuclei at each target bias potential is represented as an effective shielding potential.

### 5.1 Target Loading

The first of two possible alternative explanations considered here was the phenomenon of target loading [20], wherein the target may produce an uneven emission of neutral D atoms due to progressive heating of the target by the beam impact. Because fusion rate depends on target density as well as beam density, the observed enhancement in the fusion rate could be explained by this variation of the target density.

To address this possibility of a progressive beam loading effect, we used a stochastic sampling of the arrangement of target voltage, $V_b$ values: 0, –4, –8, –12, –16, and –20 kV. Data was collected for each $V_b$ in this randomly structured queue. The queue was re-populated with another random shuffle of this set of values after every $6^{th}$ iteration. The use of a constant set of values ensured that every $V_b$ element was equally represented in the pooled data set for statistical comparison, and the random order in which the $V_b$ values were set averaged out the effect, if any, of target loading.

Each $V_b$ was set for a period of 5 seconds (reducing the effect of long-time constant temperature effects) and neutron counts were measured. Neutron counts were then averaged for each $V_b$ value and compared with the case without bias (0 keV). The observed results were consistent with the original experimental results and showed a clear correlation between an increase in neutron yield and increasing negative biases on the target independent of the beam-loading history. We, therefore, assert that the results of this stochastic sampling experiment eliminate target loading as a significant influencing factor for the observed increase in fusion rate.

### 5.2 Beam Re-Ionization from Electron Emission

A second possible alternate explanation for the fusion rate enhancement is that the neutralized beam may have become re-ionized by electrons emitted from the biased target. Since fusion depends on the probability of the two atomic nuclei penetrating the Coulomb barrier, an increase in final relative velocity may enhance the likelihood of nuclei to fuse. Though the neutral beam would be effectively unaffected by the target bias during most of the transit through the chamber, the acceleration of the newly formed ions toward the negatively biased target would be the equivalent of a grounded target and a higher-energy beam.

The neutral particles in the beam could become ionized in our system by energetic electrons emitted from the negatively biased target through mechanisms that include thermal electron emission, secondary electrons from impact of other beam particles on the target surface, and field emission of electrons. In each case the interaction distance where reionization can occur will be very close to the target surface. And the extent of ionization depends on the electron flux, the kinetic energy of the ejected electrons, and the collision probability between the ejected electron and a neutral beam particle.

To investigate this alternate explanation, the re-ionization cross section for the incoming beam and the electron flux emitted from the target must be accurately characterized to distinguish the effect of re-ionization phenomena the fusion rate enhancement to from that of Coulomb barrier reduction. Assuming a significant degree of ionization is possible, however, we can create an upper limit case for effective increase in fusion rate due only to beam acceleration.

Considering the interaction distance of the ejected electrons to ionize the incoming beam particles to be on the order of the mean free path at the deuterium pressure in the chamber, we can predict the approximate magnitude of the effect of reionization on the rate of fusion events. For a chamber pressure of 5.3 mTorr at 300 K, the mean free path is 27 mm. For a target bias of –20 kV an ionized deuteron, initially at the beam velocity, gains only an additional 0.5 keV of kinetic energy over this distance using the velocity calculation $v^2 = v_0^2 + 2a\lambda$, where $a$ is the acceleration $a = \frac{Eq}{m}$, and $\lambda$ is the mean free path of a deuterium atom using the Van der Waals radius of the deuterium molecule.

The velocity of the accelerated ions at the time of collision with a target deuteron results in a fusion cross section calculated from the new system energy instead of the initial beam energy. The expected enhancement of fusion probability for a neutral beam particle with a speed $v_0$ that is ionized within one mean free path of an electron from the target, based on Eq. 8, and assuming the beam and target deuteron densities remain constant, is given by

$$\frac{(R-R_0)}{R_0} = \frac{(v\sigma - v_0\sigma_0)}{v_0\sigma_0} \tag{12}$$

For an initial beam energy of 25 keV, therefore, we would expect an enhancement of fusion rate of no more than 14 percent with a target bias of –20 kV if the beam reionization is complete. This is significantly less than the >60 percent enhancement illustrated in Fig. 2. We can, therefore, conclude that the enhancement of fusion that we observe with the applied bias is due primarily to a modification in the effective fusion cross section.

## 6. Conclusion

The beam-target experiments described in this report demonstrated an enhancement of neutron yields when the target is negatively biased. Results are consistent with Alpha Ring's theory of Coulomb barrier reduction caused by electron shielding. The use of a neutral beam ensures that beam particles do not gain substantial energy from the target bias, and residual ion acceleration alone is unable to create sufficient acceleration to account for observed effects under these conditions. Experimental conditions eliminate reactant densities and target loading irregularities as a possible alternate explanation for fusion rate enhancement.

**Appendix A: Wave Approach for Tunneling Probability**

The basic 1-D quantum equation can be written as (Eq. 2 in the main text):

$$\left(\frac{d^2}{dr^2} + \frac{2mE_k(r)}{\hbar^2}\right)\psi(r) = 0 \tag{A1}$$

Here $E_k(r)$ is the $r$-dependent kinetic energy and $\psi$ is the wave function, which can be represented in wave mechanics as

$$\psi(r) = e^{i\phi(r)} \tag{A2}$$

Using Eq. A2 in Eq. A1, we obtain

$$-i\frac{d^2\phi}{dr^2} + \left(\frac{d\phi}{dr}\right)^2 = \frac{2mE_k(r)}{\hbar^2} \tag{A3}$$

Assume $\phi(r)$ is a slowly varying function such that the second-order derivative in Eq. A3 is negligible. From Eq. A3, we solve for $\phi(r)$ and the wave function becomes

$$\psi(r) = e^{i\int \frac{\sqrt{2mE_k(r)}}{\hbar}dr} \tag{A4}$$

Assume the particle is originally at $r_0$ and the probability of the particle appearing at the nucleus position $r_2$ can be calculated as

$$P = \left|\frac{\psi(r_2)}{\psi(r_0)}\right|^2 = \left|e^{2i\int_{r_0}^{r_2} \frac{\sqrt{2mE_k(r)}}{\hbar}dr}\right| = \left|e^{2i\int_{r_0}^{r_2} \frac{\sqrt{2m[E-U(r)]}}{\hbar}dr}\right| \tag{A5}$$

At $r_0$, the energy $E$ is larger than the potential energy and the kinetic energy is positive. However, in the tunneling region near $r_2$, the energy is lower than the potential energy and the kinetic energy value is negative. The transition occurs at the radius $r_1$ where the kinetic energy is zero,

$$0 = E_k(r_1) = E - U(r_1) \tag{A6}$$

The integral in Eq. A5 can be divided into two regions: one for $E_k > 0$ between $r_0$ and $r_1$ and one for $E_k < 0$ between $r_1$ and $r_2$,

$$P = \left|e^{-2\int_{r_1}^{r_2} \frac{\sqrt{-2mE_k(r)}}{\hbar}dr + 2i\int_{r_0}^{r_1} \frac{\sqrt{2mE_k(r)}}{\hbar}dr}\right| \tag{A7}$$

The term with integration between $r_0$ and $r_1$ becomes one after taking the absolute value. So, the probability in Eq. A7 is reduced to

$$P = \left|e^{-2\int_{r_1}^{r_2} \frac{\sqrt{-2mE_k(r)}}{\hbar}dr}\right| = e^{-\frac{2\sqrt{2m}}{\hbar}\int_{r_1}^{r_2} \sqrt{U(r)-E}\,dr} \tag{A8}$$

The tunneling probability expressed in Eq. A8 is identical to the result of WKB approximation.

## Appendix B: Energy dependence of 3-D Fusion Cross Section

In most papers and textbooks, the fusion cross section is given as in [21–23] without providing any detailed derivation. The dependence of $E^{-1}$ outside the exponential factor was usually established with hand-waving arguments. In this appendix, we will use a full wave approach to show how we obtain the correct form of 3-D fusion cross section.

The tunneling probability given in Eq. A8 is based on the 1-D barrier. This approach is incomplete for a Coulomb potential because the barrier is basically a 3-D structure [24–30]. Most incoming projectiles interact with the target nucleus with an offset. This process is like that of Rutherford scattering, which can only be treated as 3-D interactions with the scattering potential. Based on this consideration, the 3-D tunneling cross section is best described as

$$\sigma_t = 2\pi \int_0^\pi \sigma_\theta P(\theta) \sin\theta d\theta \tag{B1}$$

Here $\sigma_t$ is the 3-D tunneling cross section, $\sigma_\theta$ is the differential scattering cross section, and $P(\theta)$ is the tunneling probability at the scattering angle of $\theta$. The differential scattering cross section from a Coulomb potential can be obtained from quantum scattering theory.

$$\sigma_\theta = \frac{m^2 k_e^2}{4\hbar^4 k^4} \sin^{-4}\frac{\theta}{2} = \frac{k_e^2}{16E^2} \sin^{-4}\frac{\theta}{2} = \frac{\rho^2}{16} \sin^{-4}\frac{\theta}{2} \tag{B2}$$

Here $k_e = \frac{e^2}{4\pi\epsilon}$ is the Coulomb constant, $E = \frac{p^2}{2m} = \frac{\hbar^2 k^2}{2m}$ is the kinetic energy of the projectile, and $\rho = \frac{k_e}{E}$ is the closest radial position of the head-on projectile.

Consider the target nucleus at the origin of the spherical coordinate. The energy of the projectile can be divided into two parts: radial and azimuthal energies,

$$E = \frac{m}{2} v^2 = \frac{m}{2}(v_r^2 + v_\theta^2) = E_r + E_\theta \tag{B3}$$

The radial tunneling probability is a 1-D process and can be taken from Appendix A as

$$P_\theta = e^{-\frac{2\sqrt{2m}}{\hbar} \int_{r_1}^{r_2} \sqrt{U(r) - E_r}\, dr} = e^{-\frac{2\sqrt{2m}}{\hbar} \int_{r_1}^{r_2} \sqrt{U(r) + E_\theta - E}\, dr} \tag{B4}$$

The azimuthal energy can be expressed in terms of angular momentum $L$,

$$E_\theta = \frac{m v_\theta^2}{2} = \frac{m^2 r^2 v_\theta^2}{2mr^2} = \frac{L^2}{2mr^2} \tag{B5}$$

Using Eq. B5 in Eq. B4, we have

$$P_\theta = e^{-\frac{2\sqrt{2m}}{\hbar} \int_{r_1}^{r_2} \sqrt{U(r) + \frac{L^2}{2mr^2} - E}\, dr} \tag{B6}$$

This is the familiar form of 1-D tunneling probability in term of the effective radial potential of $U_{eff}(r) = U(r) + \frac{L^2}{2mr^2}$. Since angular momentum is an invariant of the system, we have $L =$ $mvb$ at the infinity, where $v$ is the projectile velocity and $b$ is the offset of the projectile. Now, $P_\theta$ can be written as

$$P_\theta = e^{-\frac{2\sqrt{2m}}{\hbar}\int_{r_1}^{r_2}\sqrt{\frac{k_e}{r}+\frac{b^2}{r^2}E-E}\,dr} = e^{-\frac{2\sqrt{2mE}}{\hbar}\int_{r_1}^{r_2}\sqrt{\frac{\rho}{r}+\frac{b^2}{r^2}-1}\,dr} \tag{B7}$$

The integral in Eq. B7 is complicated but can be carried out analytically. Keeping only the lowest order terms in the limit of $r_2 \ll r_1$, we have

$$P_\theta = e^{-\sqrt{\frac{E_G}{E}}\left(1+\frac{4b}{\pi\rho}\right)} \tag{B8}$$

Using Eqs. B2 and B8 in Eq. B1, the resulting equation is

$$\sigma_t = \frac{\pi\rho^2}{8}\int_0^\pi e^{-\sqrt{\frac{E_G}{E}}\left(1+\frac{4b}{\pi\rho}\right)}\sin^{-4}\frac{\theta}{2}\sin\theta\,d\theta \tag{B9}$$

From the scattering theory, the scattering angle is related to the offset [22]

$$\tan\frac{\theta}{2} = \frac{k_e}{2Eb} = \frac{\rho}{2b} \tag{B10}$$

Using the identity $\left(\sin^{-4}\frac{\theta}{2}\sin\theta\right)d\theta \equiv -\left(4\cot\frac{\theta}{2}\right)d\left(\cot\frac{\theta}{2}\right)$ and Eq. B10 in Eq. B9, we get

$$\sigma_t = -\frac{\pi\rho^2}{2}\int_0^\pi e^{-\sqrt{\frac{E_G}{E}}\left(1+\frac{2}{\pi}\cot\frac{\theta}{2}\right)}\cot\frac{\theta}{2}\,d\left(\cot\frac{\theta}{2}\right) \tag{B11}$$

With the substitution of variable, $x = \cot\frac{\theta}{2}$, Eq. B11 becomes

$$\sigma_t = \frac{\pi\rho^2}{2}e^{-\sqrt{\frac{E_G}{E}}}\int_0^\infty e^{-2\sqrt{\frac{E_G}{E}}x}x\,dx \tag{B12}$$

The 3-D tunneling cross section is finally obtained as

$$\sigma_t = \frac{\pi\rho^2}{8}\frac{E}{E_G}e^{-\sqrt{\frac{E_G}{E}}} = \frac{\pi k_e^2}{8E_G E}e^{-\sqrt{\frac{E_G}{E}}} \propto \frac{1}{E}e^{-\sqrt{\frac{E_G}{E}}} \tag{B13}$$

Multiplying the astrophysical factor $S(E)$ to represent the fusion probability after tunneling, we have the well-known fusion cross section formula

$$\sigma_t = \frac{S(E)}{E}e^{-\sqrt{\frac{E_G}{E}}} \tag{B14}$$

Eq. B14 is the result of the quantum approach, derived from an integral of the product of Eqs. B2 and B4, both based on wave mechanics. The final form of the cross-section expression in the 3D case is identical to that derived in the 1D case.

**Appendix C: Fast Neutron Detector**

One of the neutron detection mechanisms used in the present experiment is a proton-recoil fast neutron detector (FND). The FND provides a method to directly detect the 2.45 MeV neutrons that are produced from the D-D fusion reaction [Unpublished results]. This detector contrasts with the He-3 thermal neutron detector, which relies on detecting the neutrons after being substantially slowed down for neutron capture. Fig. C-1 details the construction of the neutron detector, which consists of a plastic scintillator (EJ-276) optically coupled to a photomultiplier (PMT). The PMT is biased positive using a resistive and capacitive divider (PMT Base) whose signals are read out into an analog-to-digital (ADC) converter. A magnetic shield is placed around the device to suppress the effects of stray magnetic fields on the operation of the PMT. The typical applied voltage on the PMT is +800 V with a current draw of 0.200 mA.

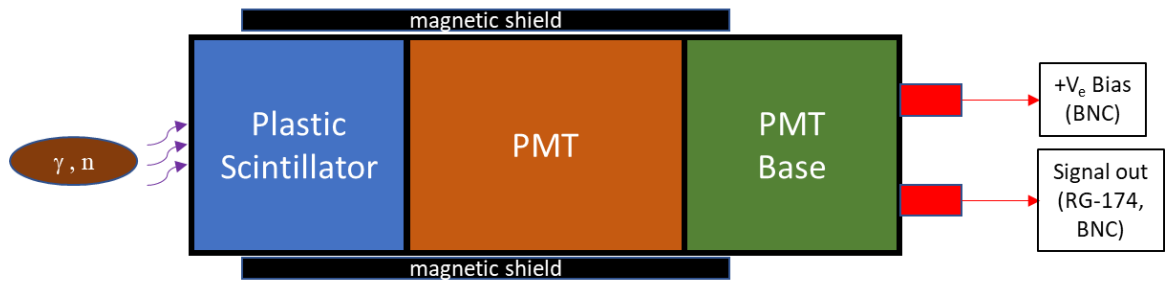

Figure C-1: Schematic of the fast neutron detector using a plastic scintillator.

A typical installation of the FND places the scintillator region at a fixed distance and orientation with respect to the neutron source. Fig. C-2 shows the FND adjacent to the beam-target assembly described in this work and illustrated in Fig. 2. The detector was in this position, near the aluminum target, for all fusion experiments reported.

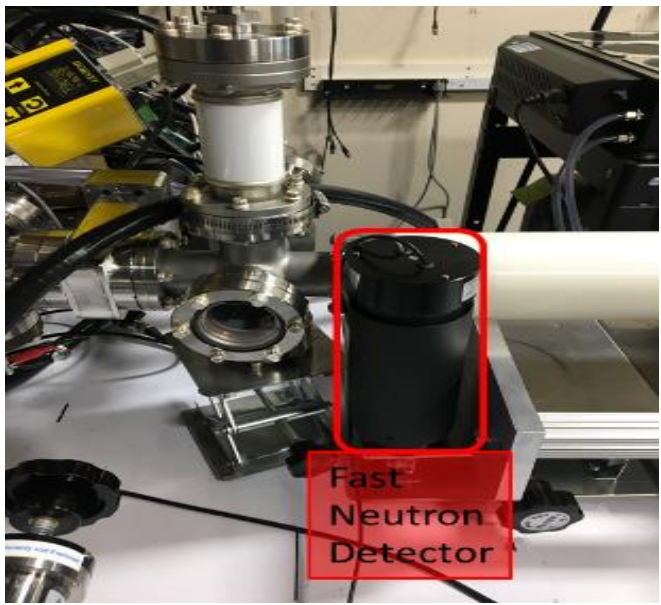

Figure C-2 Position of the fast neutron detector relative to the neutron producing target. The scintillator is located approximately 4" from the target.

The pulses generated from the fast neutron detector are discretized and further analyzed using pulse-shape-discrimination (PSD). Alpha Ring has developed a proprietary data analysis system which can provide very precise separation between the neutron and gamma signals. Fig. C-3 shows a typical neutron pulse that is observed on the oscilloscope with proper termination. The pulse width of the signals varies between 50–200 ns depending on the pulse height, which can vary between ~0.2 mV to 2 V. However, since we know the maximum energy of the neutrons produced (2.45 MeV), we can simply discard pulses with energy higher than this value.

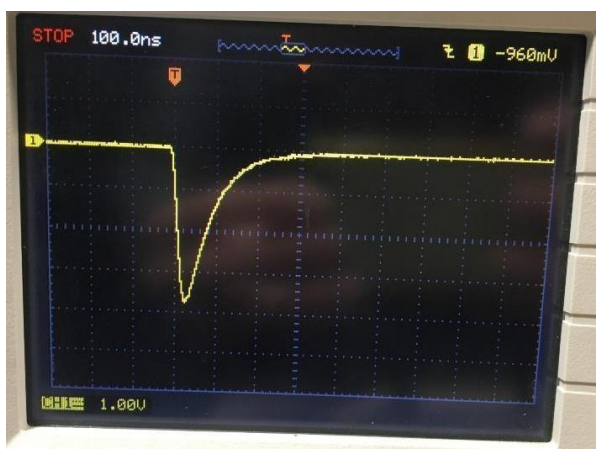

Figure C-3: A typical radiation pulse from the fast neutron detector. Pulse-shape discrimination must be used to distinguish the identity of the pulse.

A proprietary data analysis system has been developed by Alpha Ring to precisely determine the particle identity of each pulse from the fast neutron detector [Unpublished results]. Pulse-shape-discrimination is a well-established method used by the industry to accurately discriminate between neutron and gamma signals. Fig. C-4 shows the typical dataset with mixed neutron and gamma signals. We see a clear ability to separate the two clusters which represent neutrons and gammas using an optimized clustering algorithm.

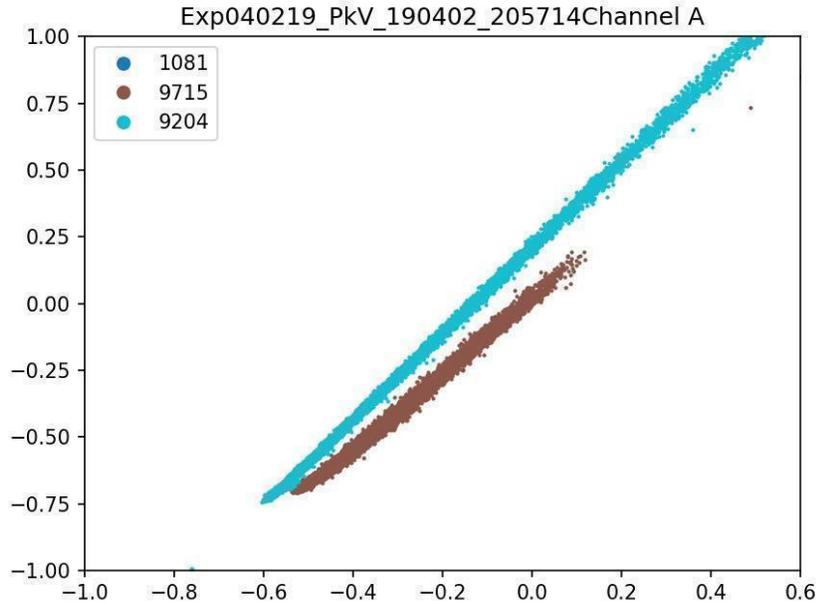

Figure C-4: Typical dataset from the experiment which contains both neutron and gamma signals. Using our proprietary clustering algorithm, gammas and neutrons can be separated into two clusters to provide quantitative results of the neutron production process. Here the brown cluster represents neutrons, and the cyan cluster represents gammas.

**Acknowledgments**

We wish to thank Tay Nguyen and Belinda Mei for their help in programming and coordination. The encouragement, support, and editing from Shahi Ghanem has been essential to the progress of this work. This research was supported by the Independent Research Fund of Alpha Ring International Limited.

**Author Contributions**

**Alfred Y. Wong**: Conceptualization, Methodology, Writing- Reviewing and editing, Resources, Supervision.  **Alexander Gunn**: Software, Validation, Formal analysis, Data curation, Writing - Reviewing and editing.  **Allan X. Chen**: Software, Investigation, Data curation, Validation, Formal analysis Project administration, Writing - Original draft.  **Chun-Ching Shih**: Formal analysis, Writing - Reviewing and editing.  **Mason J. Guffey**: Writing - Reviewing and editing.

**Funding**

Alpha Ring International Limited has been fully funded by private individual investors and a venture capital firm for research and development of its technology.

**Conflicts of Interest**

The authors declare that they have no known competing financial interests or personal relationships that could have appeared to influence the work reported in this paper.

# References

1. H. W. Becker, C. Rolfs, and H. P. Trautvetter, "Low-Energy Cross Sections for 11B(p,3α)," Z. phys. A **327**, 341 (1987)
2. C. Angulo, and S. Engstler, G. Raimann, C. Rolfs, W. H. Schulte, and E. Somorjai, "The Effect of Electron Screening and Resonances in (p,a) Reactions on 10B and 11B at Thermal Energies," Z. Phys. A **345**, 231 (1993)
3. F. C. Barker, "Electron Screening in Reactions Between Light Nuclei," Nuclear Phys. A **707**, 277 (2002)
4. J. M. Davidson, H. L. Berg, M. M. Lowry, M. R. Dwarakanath, A. J. Sierk, and P. Batay-Csorba, "Low-Energy Cross Sections for 11B(p,3α)," Nucl. Phys. A **315**, 253 (1979)
5. H. J. Assenbaum, K. Langanke, and C. Rolfs, "Effect of Electron Screening on Low-Energy Fusion Cross Sections," Z. Phys. A **327**, 461 (1987)
6. D. D. Clayton, Principles of Stellar Evolution and Nucleosynthesis, Univ. of Chicago Press (1983)
7. E. E. Salpeter, "Electron Screening and Thermonuclear Reactions," J. Phys. **7**, 373 (1954)
8. C. Bonomo, G. Fiorentini, Z. Fulop, L. Gang, G. Gyurky, K. Langanke, F. Raiola, C. Rolfs, E. Somorjai, F. Streider, J. Winter, and M. Aliotta, "Enhanced Electron Screening in d(d,p)t for Deuterated Metals: A Possible Classical Explanation," Nucl. Phys. A **719**, C37 (2003)
9. F. Raiola, L. Gang, C. Bonomo, G. Gyurky, M. Aliotta, H. W. Becker, R. Bonetti, C. Broggini, P. Corvisiero, A. D'Onofrio, Z. Fulop, G. Gervino, L. Gialanella, and M. Junker, "Enhanced Electron Screening in d(d,p)t for Deuterated Metals," Eur. Phys. J. A **19**, 283 (2004)
10. F. Raiola, B. Burchard, Z. Fulop, G. Gyurky, S. Zeng, J. Cruz, A. DiLeva, B. Limata, M. Fonseca, H. Luis, M. Aliotta, H. W. Becker, C. Broggini, A. D'Onofrio, and L. Gialanella," Electron Screening in d(d,p)t for Deuterated Metals: Temperature Effects," J. Phys. G **31**, 1141 (2005)
11. A. Krauss, H. W. Becker, H. P. Trauvetter, and C. Rolfs, "Low-Energy Fusion Cross Sections of D + D and D + 3He Reactions," Nucl. Phys. A **465**, 150 (1987); **467**, 273 (1987)
12. S. Ichimaru, "Strongly Coupled Plasmas: High-Density Classical Plasmas and Degenerate Electron Liquids," Rev. Mod. Phys. **54**, 1017 (1982)
13. S. Ichimaru and K. Utsumi, "Screening Potential and Enhancement of Thermonuclear Reaction Rate Due to Relativistic Degenerate Electrons in Dense Multi-Ionic Plasmas,' Astrophys. J. **278**, 382 (1984)
14. Y. E. Kim and A. L. Zubarev, "Theoretical Interpretation of Anomalous Enhancement of Nuclear Reaction Rates Observed at Low Energies with Metal Targets," Jap. J. Appl. Phys. **46**, 1665 (2007)

15. S. Ichimaru and H. Kitamura, Ultradense Hydrogen in Astrophysics, High-Pressure Metal Physics and Fusion Studies, in (AIP, 2010), pp. 541–550.
16. S. Ichimaru and H. Kitamura, "Pycnonuclear Reactions in Dense Astrophysical and Fusion Plasmas," Phys. Plasmas **6**, 2649 (1999)
17. A. Y. Wong and C.-C. Shih, "Approach to Nuclear Fusion Utilizing Dynamics of High-Density Electrons and Neutrals," ArXiv:1908.11068 [Physics.Plasm-Ph] (2019)

18. Qing Ji, "Compact Permanent Magnet Microwave-Driven Neutron Generator," AIP Conference Proceedings 1336, 528 (2011)
19. J. Kim and H. H. Haselton, "Analysis of Particle Species Evolution in Neutral Beam Injection Lines," Oak Ridge National Laboratory report (1978)
20. J. F. Ziegler and H. H. Anderson, *Hydrogen Stopping Powers and Ranges in All Elements*, Pergamon Press Inc. (1977)
21. L. Schiff, *Quantum Mechanics* McGraw-Hill College Sec. 21 & 34 (1968)
22. S. Atzeni, *The Physics of Inertial Fusion,* Clarendon Press Chapter 1 (2004)
23. E. G. Adelberger, et. al., "Solar Fusion Cross Sections," Rev. Mod. Phys. 70, 1265 (1998)
24. H. Yukawa, "On the Interaction of Elementary Particles," Proc. Phys. Math. Soc. Jap. **17**, 48 (1935)
25. L. Hulthen, "Über die Eigenlösungen der Schrödingergleichung des Deuterons," Ark. Mat. Astron. Fys. **28A**, 1-12 (1942)
26. S. T. Ma, "On the Coulomb and Hulthen Potentials," Aust. J. Phys. **7**, 365 (1954)
27. N. R. Arista, A. Gras-Marti, and R. A. Baragiola, "Screening Effects in Nuclear Fusion of Hydrogen Isotopes in Dense Media," Phys. Rev. A **40**, 6873 (1989)
28. D. Agboola, "Schrodinger Equation with Hulthen Potential Plus Ring-Shaped Potential," Physica Scripta **80**, 065304 (2009)
29. Wahyulianti, "Construction of Solvable Potential Partner of Generalized Hulthen Potential in D-Dimensional Schrodinger Equation," AIP Conf, Proc. **1868**, 06001 (2017)
30. R. B. Leighton, *Principles of Modern Physics*, McGraw-Hill, 1967, p. 488.